\documentclass[aps,nofootinbib,preprintnumbers,twocolumn,superscriptaddress,prb,
]{revtex4}

\usepackage{amsmath}
\usepackage{amssymb}
\usepackage{amsfonts}
\usepackage{mathtools}
\usepackage{MnSymbol}
\usepackage{color,array}
\usepackage{dsfont}
\usepackage{slashed}
\usepackage{graphicx}
\usepackage{graphics}
\usepackage{subfigure}
\usepackage{epsfig}
\usepackage{bbm,bm}
\usepackage{psfrag}
\usepackage{hyperref}
\usepackage{ulem}
\usepackage{url}

\newcommand{\be}{\begin{equation}}
\newcommand{\ee}{\end{equation}}
\newcommand{\ben}{\begin{eqnarray}}
\newcommand{\een}{\end{eqnarray}}

\newcommand{\rund}[1]{ \left( #1 \right) }
\newcommand{\eck}[1]{ \left[ #1 \right] }

\newcommand{\spitz}[1]{ \left\langle  #1 \right\rangle }
\newcommand{\abs}[1]{ \left|  #1 \right| }

\newcommand{\al}[1]{\begin{align}#1\end{align}}

\newcommand{\mb}[1]{{\mathbf{#1}}}
\newcommand{\bs}[1]{{\boldsymbol{#1}}}

\newcommand{\pr}[1]{{{#1}^{\prime}}}

\begin{document}

\title{Instabilities on graphene's honeycomb lattice with electron-phonon interactions}
\date{\today}

\author{Laura Classen}
\email{classen@thphys.uni-heidelberg.de}
\affiliation{\mbox{\it 	Institut f\"ur Theoretische Physik,
Universit\"at Heidelberg,}
\mbox{\it D-69120 Heidelberg, Germany}
}

\author{Michael M. Scherer}
\email{scherer@thphys.uni-heidelberg.de}
\affiliation{\mbox{\it 	Institut f\"ur Theoretische Physik,
Universit\"at Heidelberg,}
\mbox{\it D-69120 Heidelberg, Germany}
}

\author{Carsten Honerkamp}
\email{honerkamp@physik.rwth-aachen.de}
\affiliation{Institute for Theoretical Solid State Physics, RWTH Aachen University, Germany and JARA-FIT Fundamentals of Future Information Technology}



\begin{abstract}
We study the impact of electron-phonon interactions on the many-body instabilities of electrons on the honeycomb lattice and their interplay with repulsive local and non-local Coulomb interactions at charge neutrality. To that end, we consider in-plane optical phonon modes with wavevectors close to the $\Gamma$ point as well as to the $K, -K$ points and calculate the effective phonon-mediated electron-electron interaction by integrating out the phonon modes. 
Ordering tendencies are studied by means of a momentum-resolved functional renormalization group approach allowing for an unbiased investigation of the appearing instabilities. 
In the case of an exclusive and supercritical phonon-mediated interaction, we find a Kekul\'e and a nematic bond ordering tendency being favored over the $s$-wave superconducting state. The competition between the different phonon-induced orderings clearly shows a repulsive interaction between phonons at small and large wavevector transfers.
We further discuss the influence of phonon-mediated interactions on electronically-driven instabilities induced by onsite, nearest neighbor and next-to-nearest neighbor density-density interactions. 
We find an extension of the parameter regime of the spin density wave order going along with an increase of the critical scales where ordering occurs, and a suppression of competing orders.
\end{abstract}

\maketitle

\section{Introduction}\label{intro}

Electrons in graphene feature many unusual properties that can be captured within relatively simple theoretical frameworks based on a single-particle description of the electrons close to the Fermi level\cite{novoselov,castroneto2009}. 
The remarkable success of the interplay between experiment and single-particle theory for phenomena such as the half-integer quantum Hall effect\cite{novoselov2005a} or the Klein paradox\cite{katsnelson2006a} leads to the conclusion that electron-electron interactions in pristine graphene only play a quantitative, however, not a qualitative role. 
On the charge neutral honeycomb lattice, due to the vanishing density of states for energies close to the Fermi level, qualitative changes from interactions such as strongly correlated electronic phases can only appear beyond a critical interaction strength \cite{sorella1992,khvesh2001,herbut2006}. In this case, however, depending on the type of the interaction, possible occurrences of exotic states of matter such as Quantum Spin Hall phases\cite{raghu2008,daghofer2013} and even spin liquids are under consideration\cite{meng,sorella2012}.
Doped graphene features a non-vanishing density of states at the Fermi level which enhances the role of electronic interactions as compared to the charge neutral situation and can give rise to, possibly unconventional superconductivity\cite{doniach2007,honerkamp2008}. At least, a supercurrent in graphene has been induced\cite{heersche2007} by means of a contact of a graphene sample to superconducting electrodes. 

This raises the question under which circumstances graphene can give rise to intrinsic superconductivity. In this context, the role of electron-phonon interactions for different types of superconducting states has been investigated\cite{uchoa2007,basko2008,kopnin2008,einenkel2011} with a focus on the effects of in-plane phonons that were identified in the Raman spectra of graphene\cite{piscanec2004}. 
Further ordering patterns as, e.g., a Kekul\'e order due to the electron-phonon coupling have been considered\cite{nomura2009,kharitonov2012}.
Generally, it is a difficult task to identify the leading ordering tendency given a large variety of possible ordering patterns, in particular when various interaction effects compete.

In this work, we investigate the ordering tendencies of electrons on the honeycomb lattice when electron-phonon mediated electronic interactions from in-plane optical phonons as well as short-ranged Coulomb interactions are present. Therefore, we employ a functional renormalization group approach in a momentum-resolved patching scheme for the vertex function. This method provides an unbiased investigation of the appearing instabilities and has been proven to be a reliable tool for the study of a large range of two-dimensional solid state systems with strongly-correlated phases, e.g. high-$T_c$ superconductors, such as cuprates, pnictides and has been estabilished before for investigations on the honeycomb lattice, see Refs.~\onlinecite{metzner2012,thomale2013} for recent reviews. 

This paper is organized as follows. In Sec.~\ref{model}, we introduce our model in terms of a tight-binding Hamiltonian with nearest-neighbor hopping and density-density interactions. Phonon-modes are included upon expansion of the hopping amplitude in the displacements and integrated out to give a contribution to the electron-electron interaction. In Sec.~\ref{frg}, we describe the functional renormalization group (fRG) method and discuss the $N$-patch scheme as well as the employed approximations. We present results on the ordering tendencies in Sec.~\ref{phasediag}, first by discussing exclusive in-plane optical phonons to analyze their isolated effect, see Sec.~\ref{phononia}. Depending on the ratio of the strengths of the electron-phonon couplings for phonon modes close to $\Gamma$ or $K,-K$, we find that the leading instability is of Kekul\'e or nematic type.
The interplay with short-ranged Coulomb interactions is studied and the impact of the electron-phonon coupling is discussed in Sec.~\ref{coulomb}. We draw conclusions in Sec.~\ref{conclusions}.

\section{Model Hamiltonian}\label{model}

We consider a tight-binding model of electrons on the bipartite two-dimensional honeycomb lattice with nearest-neighbor hopping 
\al{
H=-t\sum_{\spitz{i,j},s}\left(c_{A,i,s}^{\dagger}c_{B,j,s}+\text{h.c.}\right),
}
where $c_{A,i,s}^{(\dagger)}$ annihilates (creates) an electron in unit cell $i$ on sublattice $A$ with spin $s$ and analogous for sublattice $B$. 
The first sum includes all neighboring sites denoted by $\spitz{i,j}$. 
They are connected by the nearest-neighbor hopping amplitude which in graphene has been estimated to be $t\approx2.8\,$eV. 
After Fourier transformation with $c_{o,i,s}=\sum_{\mb{k}}\exp(i\mb{k}\cdot\mb{r}_i)c_{o,\mb{k},s}/\sqrt{N}$ and $o\in \{A,B\}$, the tight-binding Hamiltonian reads
\al{
H=-t\sum_{\mb{k},s}\rund{\Delta_{\mb{k}}c_{A,\mb{k},s}^{\dagger}c_{B,\mb{k},s}+\text{h.c.}}
}
with $\Delta_{\mb{k}}=\sum_{i}\exp(-i\mb{k}\cdot\bs{a}_i)$, where $\bs{a}_i$ labels the primitive lattice vectors together with zero, i.e. $i \in \{1,2,3\}$. 
Explicitly, the $\bs{a}_i$ are given by 
$\bs{a}_1= \bs{0}$, $\bs{a}_2=\sqrt{3}a \bs{e}_x$ and $\bs{a}_3= \frac{\sqrt{3}a}{2}\bs{e}_x+\frac{3a}{2}\bs{e}_y$,
where $a$ is the lattice constant. 
Diagonalization of $H$ gives two bands with two inequivalent, linear band crossing points at the Brillouin zone corners, the Dirac cones at $\mb{K}$ and $-\mb{K}$. With the spin resolved density operator $n_{i,s}=c_{o,i,s}^{\dagger}c_{o,i,s}$ we account for repulsive onsite, nearest and next-nearest neighbor interactions
\begin{equation}\label{eqn:interactions}
  H_{I}=U\sum_{i}n_{i,\uparrow}n_{i,\downarrow}+V_1\hspace{-0.1cm}\sum_{\substack{\spitz{i,j},\\ s,\pr{s}}}n_{i,s}n_{j,\pr{s}}  +V_2\hspace{-0.1cm}\sum_{\substack{\spitz{\spitz{i,j}},\\ s,\pr{s}}}n_{i,s}n_{j,\pr{s}}.
\end{equation}
An estimate for the interaction parameters can be obtained from constrained random phase approximations\cite{wehling2011}. 
Diagonalizing the single-particle Hamiltonian provides an orbital makeup for the interaction terms, i.e.~a momentum-dependent vertex in the band representation $V\rightarrow V(\mb{k}_1,n_1,\mb{k}_2,n_2,\mb{k}_3,n_3,n_4)$ determined by four band indices $n_i$, and three independent momenta $\mb{k}_i$.

\subsection{Inclusion of phonon modes}\label{phonons}

To determine the coupling of electrons and lattice displacements, we expand the hopping amplitude in the displacement fields $\mb{u}$, based on the assumption that it depends on the distance between neighboring sites, i.e. 
\begin{equation}\label{eq:displ}
t\rightarrow t-\alpha_{||}\rund{\mb{u}_i-\mb{u}_j}\cdot\bs{\hat{\delta}_{ij}}. 
\end{equation}
The expansion depends on the bond direction $\bs{\hat{\delta}_{ij}}$ pointing along one of the three nearest neighbor vectors. 
The expansion parameter is determined by ab initio calculations to be $\abs{\alpha_{||}}\approx4.4\text{eV/\AA}-5.3\text{eV/\AA}$ with $\alpha_{||}<0$\cite{piscanec2004,lazzeri2008,cappelluti2012}. 
We introduce the phonons by the usual quantization of the Fourier transformed displacement fields using the explicit expressions $\mb{u}_i=\sum_{\mb{q}}\exp(i\mb{q}\cdot\mb{r}_i)\mb{u}_{\mb{q},A}/\sqrt{N}$ and $\mb{u}_j=\sum_{\mb{q}}\exp(i\mb{q}\cdot(\mb{r}_i-\bs{a}_j))\mb{u}_{\mb{q},B}/\sqrt{N}$ for site $i$ in sublattice $A$ and its nearest neighbor $j$ in sublattice $B$. Further, 
\begin{equation}
\mb{u}_{\mb{q},o}=\sum_{\lambda}u_{\mb{q}}^{\lambda}\mb{e}_{\mb{q},o}^{\lambda}=\sum_{\lambda}\frac{1}{\sqrt{2M\Omega_{\mb{q}}^{\lambda}}}\rund{p_{\lambda,\mb{q}}+p_{\lambda,\mb{-q}}^{\dagger}}\mb{e}_{\mb{q},o}^{\lambda}\,.
\end{equation}
The carbon mass is denoted by
$M$ and $p_{\lambda,\mb{q}}^{(\dagger)}$ is the annihilation (creation) operator of a phonon in mode $\lambda$ with momentum $\mb{q}$. 
Corresponding dispersions and polarizations are given by $\Omega_{\mb{q}}^{\lambda}$ and $\mb{e}_{\mb{q},o}^{\lambda}$. 
This inclusion of lattice distortions in terms of the phonon operators leads to the following electron-phonon coupling in orbital momentum space
\al{\label{eq:eph_coupling}
	\delta H&=\frac{\alpha_{||}}{\sqrt{N}}\sum_{\substack{\mb{k},\mb{q},\\ s,\lambda}}\frac{1}{\sqrt{2M\Omega_{\mb{q}}^{\lambda}}}\notag \\ &\quad\times\eck{g_{\mb{k}}^{\lambda}(\mb{q})c_{A,\mb{k},s}^{\dagger}c_{B,\mb{k-q},s}\rund{p_{\lambda,\mb{q}}+p_{\lambda,\mb{-q}}^{\dagger}}+\text{h.c.}},
}
where $g_{\mb{k}}^{\lambda}(\mb{q})=\sum_{i}\rund{e^{i\mb{q}\cdot\bs{a}_i}\mb{e}_{\mb{q},A}^{\lambda}-\mb{e}_{\mb{q},B}^{\lambda}}\cdot\bs{\hat{\delta}}_ie^{-i\mb{k}\cdot\bs{a}_i}$. Similar couplings were obtained for Carbon nanotubes\cite{jiang2005} and by a fit to ab-initio values in graphene\cite{piscanec2004}.

Integrating out the phonons in the functional integral representation gives an effective electronic interaction
\al{
  H_\mathrm{ph-med}&=-\frac{\alpha_{||}^2}{2MN}\sum_{q,\lambda}\sum_{\substack{k,s,\\ \pr{k},\pr{s}}}\frac{1}{q_0^2+\Omega_{\mb{q}}^{\lambda 2}}\notag \\
   &\hspace{-0.3cm}\times\left[ g_{\mb{k}}^{\lambda}(\mb{q})g_{-\mb{\pr{k}}}^{\lambda}(\mb{q})^*c_{A,k,s}^{\dagger}c_{A,\pr{k},\pr{s}}^{\dagger}c_{B,\pr{k}+q,\pr{s}}c_{B,k-q,s}\right. \notag \\ &\left. +  g_{\mb{k}}^{\lambda}(\mb{q})g_{\mb{\pr{k}}}^{\lambda}(\mb{q})^*c_{A,k,s}^{\dagger}c_{B,\pr{k}-q,\pr{s}}^{\dagger}c_{A,\pr{k},\pr{s}}c_{B,k-q,s}\right.
   \notag \\
   &+ \left.\text{c.c.} \label{effint} \right],
}
mediated by the phonons. The multiindex $k=(k_0,\mb{k})$ collects the fermionic or bosonic Matsubara frequency and the wave vector. 

In the following, we will need expressions for the phonon dispersion and polarization. In principle, a calculation of the phonon spectrum  would give eigenvectors with $x$- and $y$-components of the displacements for a given lattice site that vary with the phonon-wavector $\mb{q}$. In this two dimensional, bipartite system, the eigenvectors correspond to four possible polarizations $\lambda$ being orthogonal to each other. 
In DFT calculations it was shown that the optical modes with wavevector close to $\bs{\Gamma}$ and the highest-energy modes close to $\mb{K},\mb{-K}$ give the main contributions to the electron-phonon coupling strength\cite{park2008}. This is why, we will concentrate on them and use only their energy and polarization in the phonon-mediated interaction, see Sec.~\ref{phasediag}.
This approximation simplifies the study a lot, but should not affect the qualitative results, as the smaller variation of the phonon energy due to the dispersion of the opctial modes in the denominator of Eq. (\ref{effint}) does not have a strong impact on the effective interaction.

\section{FRG method}\label{frg}

We use the functional renormalization group approach to describe the evolution from the bare action to an effective action at low energy as a function of the energy scale in an unbiased way, i.e. the structure of the effective low energy theory is not anticipated. 
This method accounts for effects beyond mean field and RPA as it also includes the interplay between different ordering tendencies. 
We use the fRG approach for the one-particle-irreducible (1PI) vertices with a momentum cutoff. For a recent review see Ref.~\onlinecite{metzner2012}. 
The 1PI vertices are generated by the effective action $\Gamma$, which is the Legendre transform of the generating functional for the connected Green's functions. 
In the effective action, the bare propagator of the system $G_0$ is modified by an infrared regulator $C^{\Lambda}$
\al{
  G_0\rund{\omega,\mb{k},b}\rightarrow G_0^{\Lambda}\rund{\omega,\mb{k},n}=\frac{C^{\Lambda}\rund{\xi_{n,\mb{k}}}}{i\omega-\xi_{n,\mb{k}}}.
}
with the single-particle energy $\xi_{n,\mb{k}}=\epsilon_{n,\mb{k}}-\mu$ in band $n$ and the chemical potential $\mu$. 
The regulator is chosen as a step function, which suppresses the modes with energy less than $\Lambda$ and reads $C^{\Lambda}\rund{\xi_{n,\mb{k}}}=\Theta\rund{\abs{\xi_{n,\mb{k}}}-\Lambda}$. 
For numerical stability, we have slightly softened the step function in the actual implementation. 
With this modification of the bare propagator, we obtain a scale-dependent effective action $\Gamma_\Lambda$ and a variation of the scale $\Lambda$ provides a renormalization group (RG) flow. 
Integration of the RG flow from an initial scale $\Lambda_0$  (typically of the order of the bandwidth) down to low energies $\Lambda\rightarrow 0$ provides a smooth interpolation between the bare and the effective action of the system.
During the lowering of the energy scale, some components of the effective two-particle interaction $V^{\Lambda}$ typically grow large and diverge at a critical scale $\Lambda_c>0$ indicating an instability towards an ordered state. The pronounced momentum structure of the near-critical interaction vertex then allows to extract an effective Hamiltonian for the low-energy degrees of freedom and determines the leading order parameter.

In the following, several approximations are employed to numerically integrate the resulting RG flow equations for the 1PI vertices efficiently:
We truncate after the two-particle interaction vertex $V^{\Lambda}$, which means that the results are of second order in the interaction. 
The flow of $V^{\Lambda}$ provides the essential information about the leading instabilities and we further neglect the effect of self-energy corrections as they couple to the flow of the interaction vertex only at third order, see e.g. Ref.~\onlinecite{SalmhoferHonerkamp}.
In addition, we do not account for the frequency dependence of the vertex $V^{\Lambda}$ and set the external frequencies to zero to single out the most singular contribution of the flow for the determination of the ground state properties of the system. This strategy has proven to provide reliable results in a large number of different two-dimensional fermionic solid state systems, cf. Refs.~\onlinecite{metzner2012,thomale2013}.
Despite these approximations, one obtains an infinite-order summation of second order diagrams which, importantly, accounts for the competition between different channels.

The flow equation for the coupling function $V^{\Lambda}\rund{k_1,k_2,k_3,n_4}$ reads
\al{
  \frac{d}{d\Lambda}V^{\Lambda}\rund{k_1,k_2,k_3,n_4}=\tau^{\Lambda}_{pp}+\tau^{\Lambda}_{ph,d}+\tau^{\Lambda}_{ph,cr}\,
}
with $k_i=(\omega_i,\mb{k}_i,n_i)$ labeling the Matsubara frequencies, the wave vectors and the bands. The particle-particle channel is given as
\al{
  \tau^{\Lambda}_{pp}=\sumint V^{\Lambda}\rund{k_1,k_2,k,\pr{n}}L^{\Lambda}\rund{k,q_{pp}}V^{\Lambda}\nonumber\rund{k,q_{pp},k_3,n_4},
}
and $\sumint=-A_{BZ}^{-1}T\sum_{\omega}\int d^2k\sum_{n,\pr{n}}$. The direct and the crossed particle-hole channels are given by
\al{
  \tau^{\Lambda}_{ph,d}&=\sumint\left[-2V^{\Lambda}\rund{k_1,k,k_3,\pr{n}}L^{\Lambda}\rund{k,q_{d}}V^{\Lambda}\rund{q_{d},k_2,k,n_4} \nonumber\right.\\
  &\quad+  \left. V^{\Lambda}\rund{k,k_1,k_3,\pr{n}}L^{\Lambda}\rund{k,q_{d}}V^{\Lambda}\rund{q_{d},k_2,k,n_4}\nonumber \right.\\
  &\quad+ \left.V^{\Lambda}\rund{k_1,k,k_3,\pr{n}}L^{\Lambda}\rund{k,q_{d}}V^{\Lambda}\rund{k_2,q_{d},k,n_4}\nonumber\right]\,,}
  and
  \al{
   \tau^{\Lambda}_{ph,cr}&=\sumint V^{\Lambda}\rund{k,k_2,k_3,\pr{n}}L^{\Lambda}\rund{k,q_{cr}}V^{\Lambda}\rund{k_1,q_{cr},k,n_4}\,,\nonumber
}
respectively, and we define the wave vectors $\mb{q}_{pp}=-\mb{k+k_1+k_2}$, $\mb{q}_{d}=\mb{k+k_1-k_3}$ and $\mb{q}_{cr}=\mb{k+k_2-k_3}$. $A_{BZ}$ denotes the area of the first Brillouin zone. The loop kernel reads
{\al{
  L^{\Lambda}\rund{k,\pr{k}}=\frac{d}{d\Lambda}\eck{G_0^{\Lambda}(k)G_0^{\Lambda}(\pr{k})}\,,
}}
with the free propagator $G_0^{\Lambda}$ due to the neglect of the self-energy.
\begin{figure}[t]
  \centering
  \includegraphics[height=0.42\columnwidth]{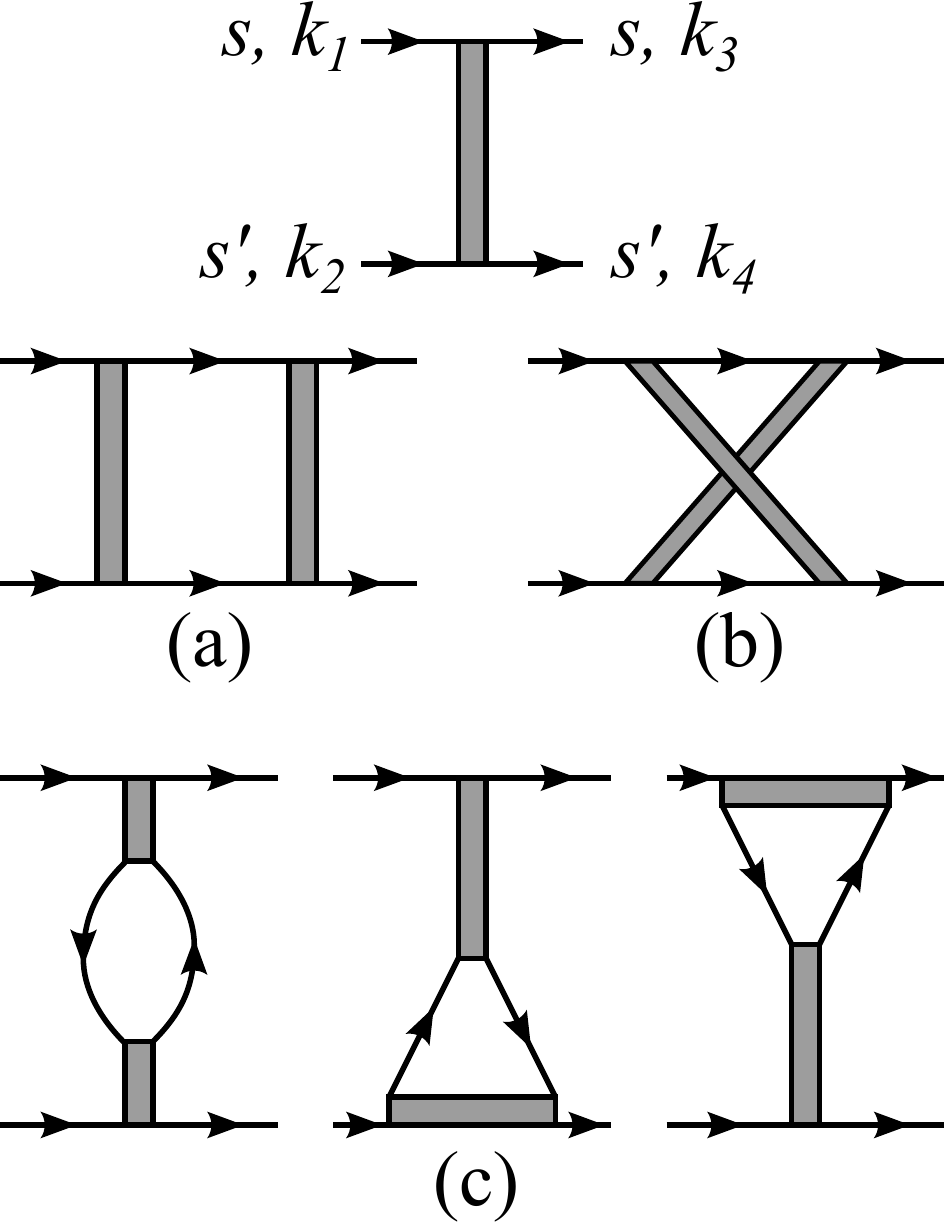}
  \hspace{0.7cm}
  \includegraphics[height=0.42\columnwidth]{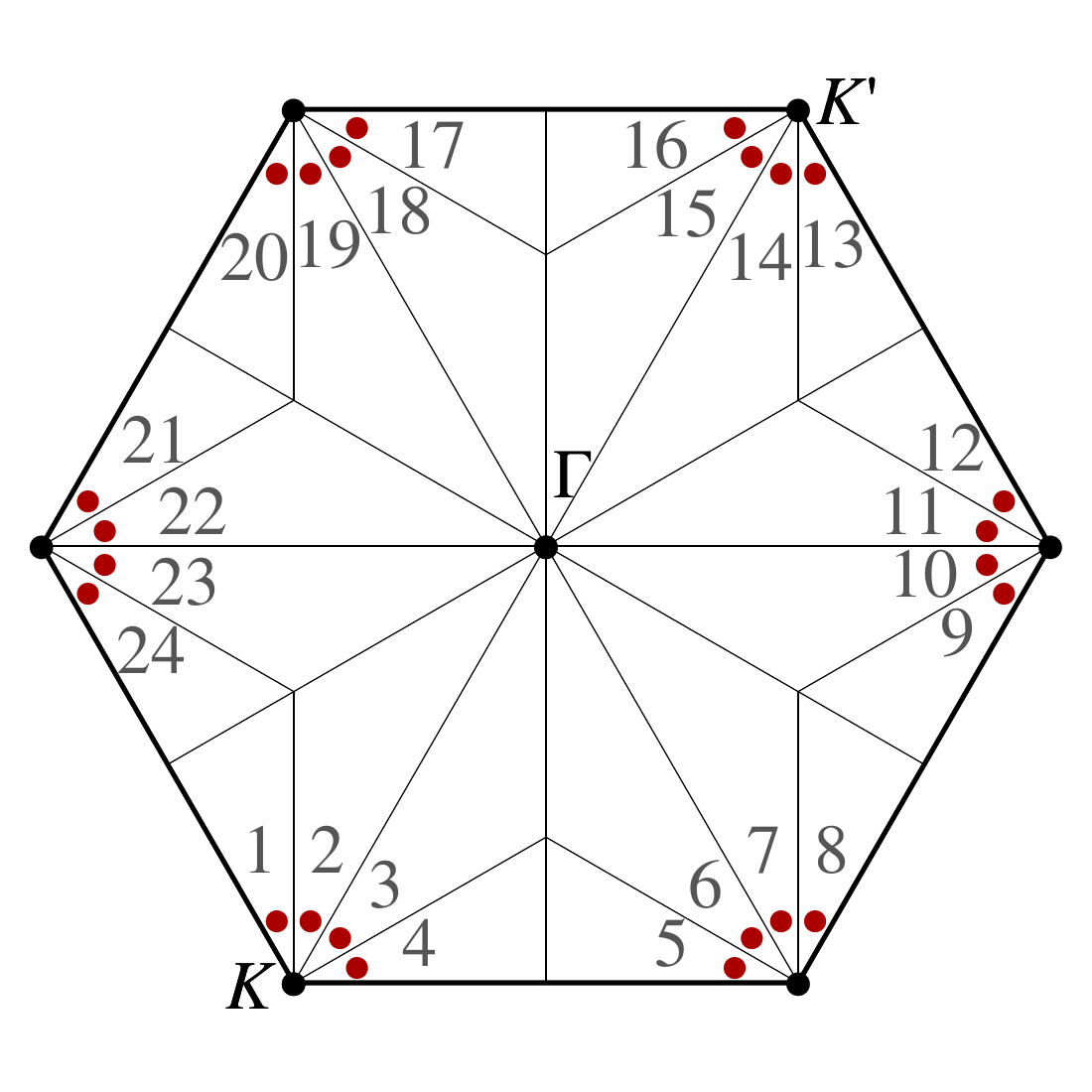}
  \caption{Left panel: Interaction vertex with spin convention (diagram on top). Below, we depict the loop contributions coming from the particle-particle channel (a), the crossed particle-hole channel (b) and the direct particle-hole channel (c). Right panel: Discretization of the momentum dependence (patching) in the Brillouin zone. Each patch is represented by a wave vector near the Fermi level as indicated by the dots.}\label{fig:patching}
\end{figure}

To solve the flow equations numerically, the wave vector dependence of the interaction vertex is discretized with a patching scheme that divides the Brillouin zone into $N$ patches, in which the vertex is approximated to be constant. 
This procedure has been established in Refs.~\onlinecite{SalmhoferHonerkamp} and successfully applied in a large number of works, cf. Ref.~\onlinecite{metzner2012}. 
The representative momentum for each patch is placed close to the Fermi level. Thereby the angular dependence of the interaction on the wavevectors is taken into account. We choose $N=24$ or $N=48$ patches as shown in Fig.~\ref{fig:patching} to check that our findings do not depend on this choice.
In the actual numerical evaluation of a diverging interaction vertex, we stop the flow at a scale $\Lambda_{\mathrm{IR}}$ where the largest component of $V_\Lambda$ is of the order of ten times the bandwith and use this $\Lambda_{\mathrm{IR}}$ as an estimate for the critical scale $\Lambda_c$.
The flow to strong coupling signals that the self-energy negligence is no longer valid. Self-energy corrections would alter the low-energy spectrum, e.g. by the appearance of a gap, and without these corrections, an emerging order appears as a divergence. 
The critical scale $\Lambda_c$ can be interpreted as an estimate for the temperature below which ordering occurs or correlations of the order parameter become important.

\section{Instabilities and phase diagram}\label{phasediag}

In this section, we analyze the phases of the honeycomb lattice system at temperature $T=0$ as a function of the bare interaction parameters $U,\ V_1$ and $V_2$ and the electron-phonon couplings $\alpha_N,\ \alpha_K$ that determine the impact of the phonons with wavevector close to $\Gamma$ and $\mb{K},-\mb{K}$, respectively.

\subsection{Purely phonon-mediated interaction}\label{phononia}

We start with the study of the isolated effect of the phonon-mediated electron-electron coupling. As mentioned in the beginning, we focus on the phonons at $\bs{\Gamma}$ and $\mb{K},-\mb{K}$. This means that we set $\Omega_{\mb{q}}^{\lambda}=\Omega_{\bs{\Gamma}}=\text{const}$ if $\mb{q}$ is close to $\bs{\Gamma}$ and $\lambda$ corresponds to the optical branches. For $\mb{q}$ in the vicinity of the Dirac points and $\lambda$ labeling the three highest-energy phonons $\Omega_{\mb{q}}^{\lambda}=\Omega_{\mb{K}}=\Omega_{-\mb{K}}=\text{const}$. The analogous approximation is used for the polarization vectors $\mb{e}_{\mb{q},a},\mb{e}_{\mb{q},b}$ and all other modes are neglected.
This ansatz accounts for the phonons that have been identified in the Raman spectrum of graphene and in DFT calculations to give the strongest electron-phonon coupling \cite{park2008,piscanec2004}. They are often referred to as $E_2$ and $A_1^{\prime}$ or $A_1,B_1$ phonons, respectively. It has been shown in Refs. \onlinecite{kharitonov2012,nomura2009} that the latter modes can give rise to a an instability corresponding to Kekul\'e ordering. We also find this to be the dominating instability for equally strong coupling of both modes, whereas for an enhanced coupling of the $E_2$ phonons, a nematic bond order is induced.

With these preliminaries and comments, we choose to parametrize the phonon-mediated contribution to the electron-electron interaction as
\begin{widetext}
\al{\label{eqn:ph_med_ia}
H_{\mathrm{ph-med}}&=-\frac{1}{N}\sum_{\substack{\mb{k_1,k_2,k_3}\\s,\pr{s}}}\left[V^{AABB}_{\mb{k_1,k_2,k_3}}c^{\dagger}_{A,\mb{k_3},s}c^{\dagger}_{A,\mb{k_4},\pr{s}}c_{B,\mb{k_2},\pr{s}}c_{B,\mb{k_1},s}+V^{ABAB}_{\mb{k_1,k_2,k_3}}c^{\dagger}_{A,\mb{k_3},s}c^{\dagger}_{B,\mb{k_4},\pr{s}}c_{A,\mb{k_2},\pr{s}}c_{B,\mb{k_1},s}+\text{h.c.}\right] \\
V^{AABB}_{\mb{k_1,k_2,k_3}}&= \left\{\begin{matrix*}[l]\alpha_{N}\sum_{\lambda=1}^2\frac{1}{\{\Omega_{\bs{\Gamma}}^{\lambda}\}^2}g_{\mb{k_3}}^{\lambda,\bs{\Gamma}}(\mb{k_3-k_1}) g_{\mb{-k_4}}^{\lambda,\bs{\Gamma}}(\mb{k_3-k_1})^* && :&&\mb{k_3-k_1} \text{ close to } \bs{\Gamma}\\  \alpha_{K}\sum_{\lambda=1}^3\frac{1}{\{\Omega_{\pm\mb{K}}^{\lambda}\}^2}g_{\mb{k_3}}^{\lambda,\pm\mb{K}}(\mb{k_3-k_1}) g_{\mb{-k_4}}^{\lambda,\pm\mb{K}}(\mb{k_3-k_1})^* && :&&\mb{k_3-k_1} \text{ close to } \pm\mb{K} \end{matrix*}\right.\\ \\
V^{ABAB}_{\mb{k_1,k_2,k_3}}&= \left\{\begin{matrix*}[l]\alpha_{N}\sum_{\lambda=1}^2\frac{1}{\{\Omega_{\bs{\Gamma}}^{\lambda}\}^2}g_{\mb{k_3}}^{\lambda,\bs{\Gamma}}(\mb{k_3-k_1}) g_{\mb{k_2}}^{\lambda,\bs{\Gamma}}(\mb{k_3-k_1})^* && :&&\mb{k_3-k_1} \text{ close to } \bs{\Gamma}\\  \alpha_{K}\sum_{\lambda=1}^3\frac{1}{\{\Omega_{\pm\mb{K}}^{\lambda}\}^2}g_{\mb{k_3}}^{\lambda,\pm\mb{K}}(\mb{k_3-k_1}) g_{\mb{k_2}}^{\lambda,\pm\mb{K}}(\mb{k_3-k_1})^* && :&&\mb{k_3-k_1} \text{ close to } \pm\mb{K} \end{matrix*}\right.
}
\end{widetext}
with $\mb{k_4}=\mb{k_1+k_2-k_3}$ due to momentum conservation and fixed polarization in $g_{\mb{k}}^{\lambda,\mb{Q}}(\mb{q})=\sum_{i}\rund{e^{i\mb{q}\cdot\bs{a}_i}\mb{e}_{\mb{Q},A}^{\lambda}-\mb{e}_{\mb{Q},B}^{\lambda}}\cdot\bs{\hat{\delta}}_ie^{-i\mb{k}\cdot\bs{a}_i}$ as motivated before. $\{\Omega_{\mb{Q}}^{\lambda}\}$ labels the value of the phonon energy, so that the interaction parameter $\alpha_{N/K}$ has units of energy\footnote{$\alpha_N$ and $\alpha_K$ should not be confused with $\alpha_{||}$ from the expansion of the hopping amplitude $\alpha_{N/K}=\alpha_{||}^2/(2M[\Omega_{\bs{\Gamma}/\mb{K}}^{\lambda}]^2)$. With the ab-initio values for $\alpha_{||}$ from above, the interaction parameter is of the order $\alpha_{N/K}\approx0.12t-0.17t$.}.
\begin{figure}[t!]
\centering
\subfigure[Bare phonon-mediated interaction from Eq. (\ref{eqn:ph_med_ia}).\label{fig:start}]{\includegraphics[height=.3\columnwidth]{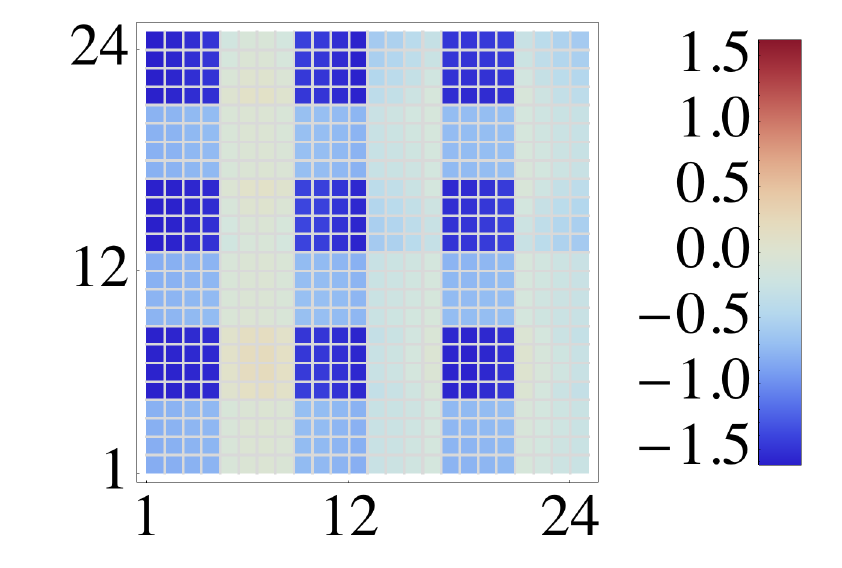}
\hspace{0.05cm}
\includegraphics[height=.3\columnwidth]{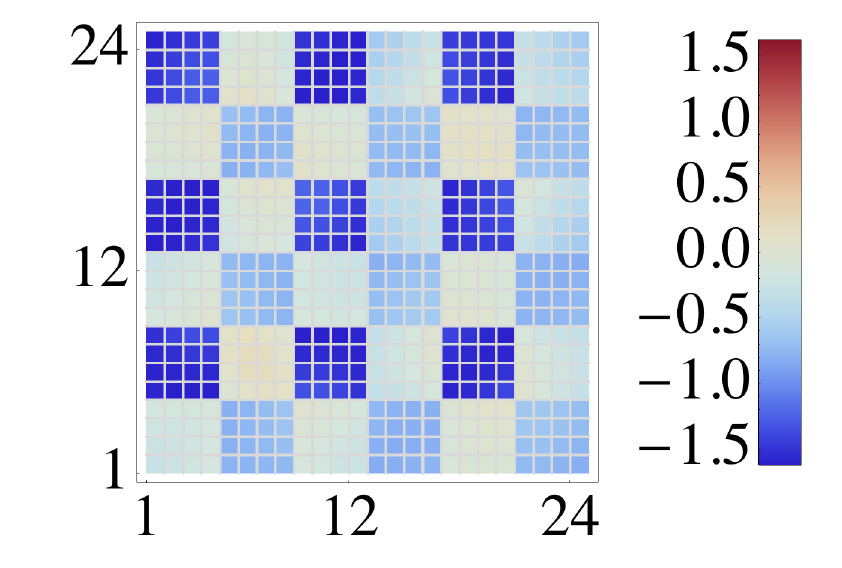}}
\\
\subfigure[Effective low-energy interaction near the critical scale for dominating $\alpha_K$.\label{fig:eff_phmed_K}]
{\includegraphics[height=.3\columnwidth]{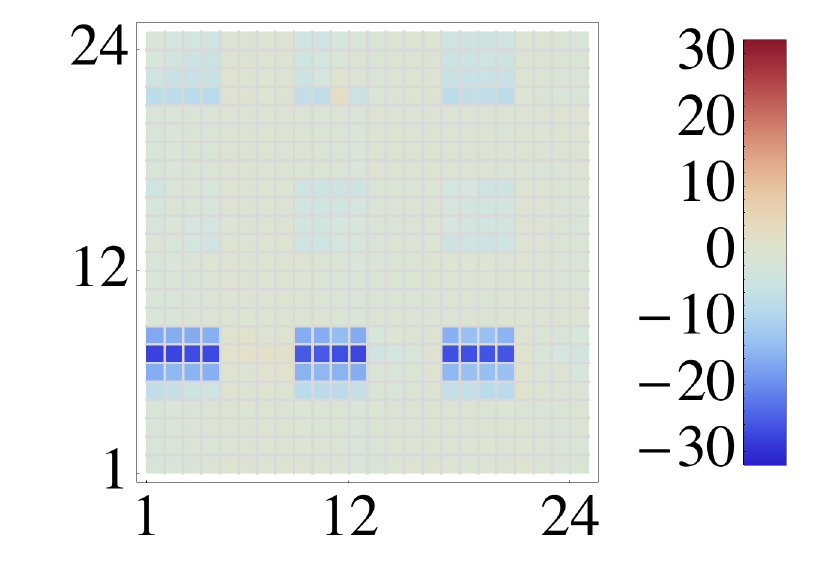}
\hspace{0.05cm}
\includegraphics[height=.3\columnwidth]{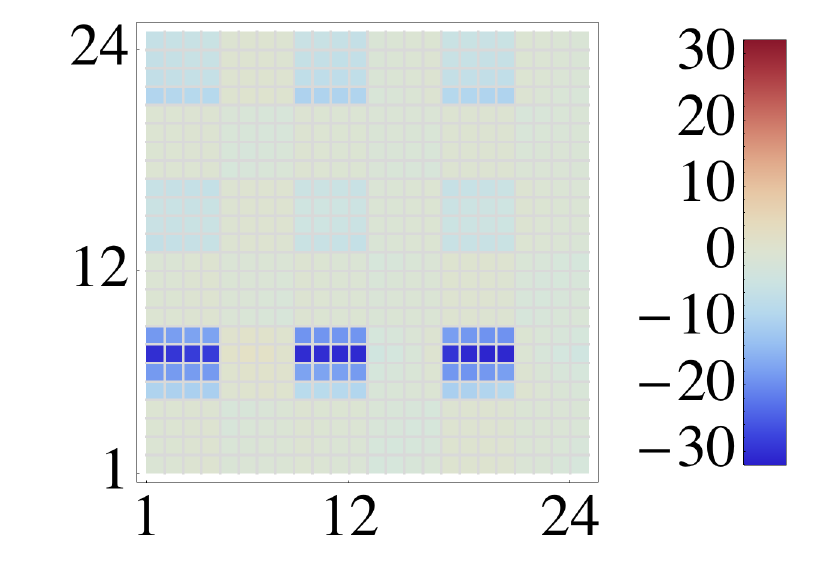}}
\\
\subfigure[Effective low-energy interaction near the critical scale for dominating $\alpha_N$.\label{fig:eff_phmed_N}]
{\includegraphics[height=.3\columnwidth]{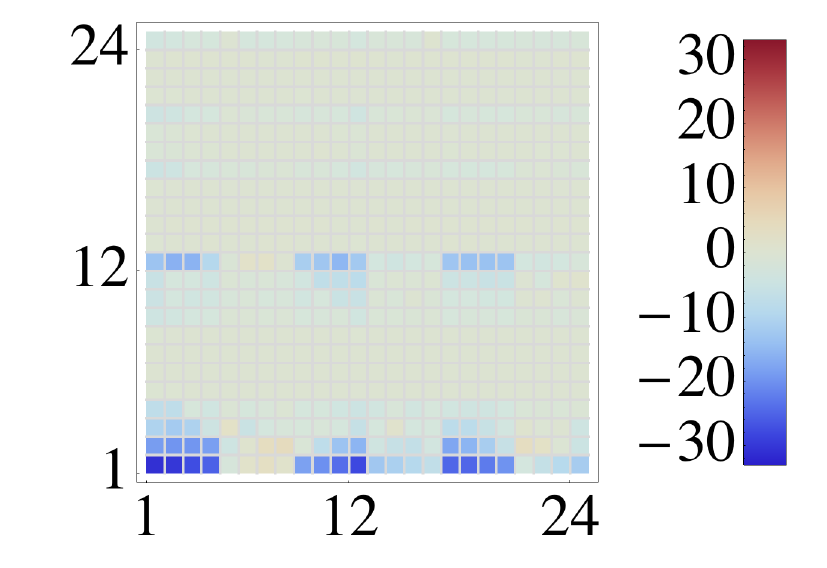}
\hspace{0.05cm}
\includegraphics[height=.3\columnwidth]{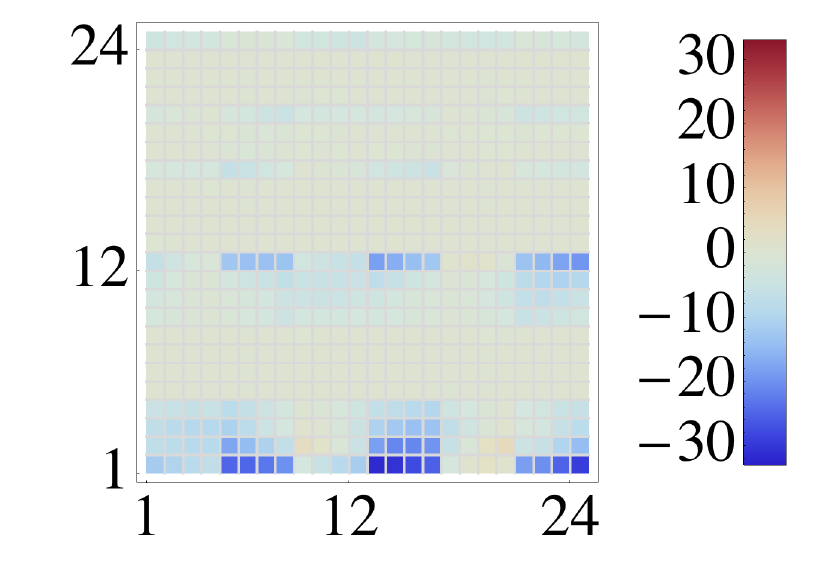}}
\caption{Bare and effective phonon-mediated interaction in units of the hopping amplitude $t$. The numbers on the axes correspond to the patches of Fig.~\ref{fig:patching}. The horizontal axes shows $\mb{k}_2$, the vertical $\mb{k}_1$ and $\mb{k}_3$ is fixed on the first patch. Orbital combinations are for all subfigures (a), (b) and (c), $o_1=o_4\neq o_2=o_3$ (left panel) and $o_1=o_2\neq o_3=o_4$ (right panel).}
\end{figure}
As is well known and also discussed in the context of RG e.g. in Ref.~\onlinecite{fu2006}, the phonon-mediated interaction is suppressed for frequencies larger than the phonon frequency $\Omega$. Resolving this frequency dependence of the interactions in the RG flow with frequency-independent interactions requires some physical insights, at least if one wants to reduce the numerical effort\footnote{Note that in some recent fRG works \cite{honerkampfulee2007,uebelackerhonerkamp2012,giering2012}, the frequency dependence of the interactions have been taken into account.}. Usually one tries to replace the frequency dependence with a dependence on the electronic excitation energy. One reasonable choice for studying the phonon-mediated interaction case separately would then be to
only include interactions of electrons with excitation energies below $\Omega$. This would correspond to start the RG flow only at RG scale $\Lambda_0=\Omega$. In this work, however, we chose to start the flow already at the bandwidth $\Lambda_0=3t$. This can be viewed as ignoring the retardation and artificially enhances the impact of electron-phonon interactions and makes its effects clearly visible. 

\begin{figure}[t!]
\centering
\subfigure[Critical scale vs. $\alpha_N$ or $\alpha_K$ for different phonon scenarios.\label{fig:alphacrit}]{\includegraphics[height=.48\columnwidth]{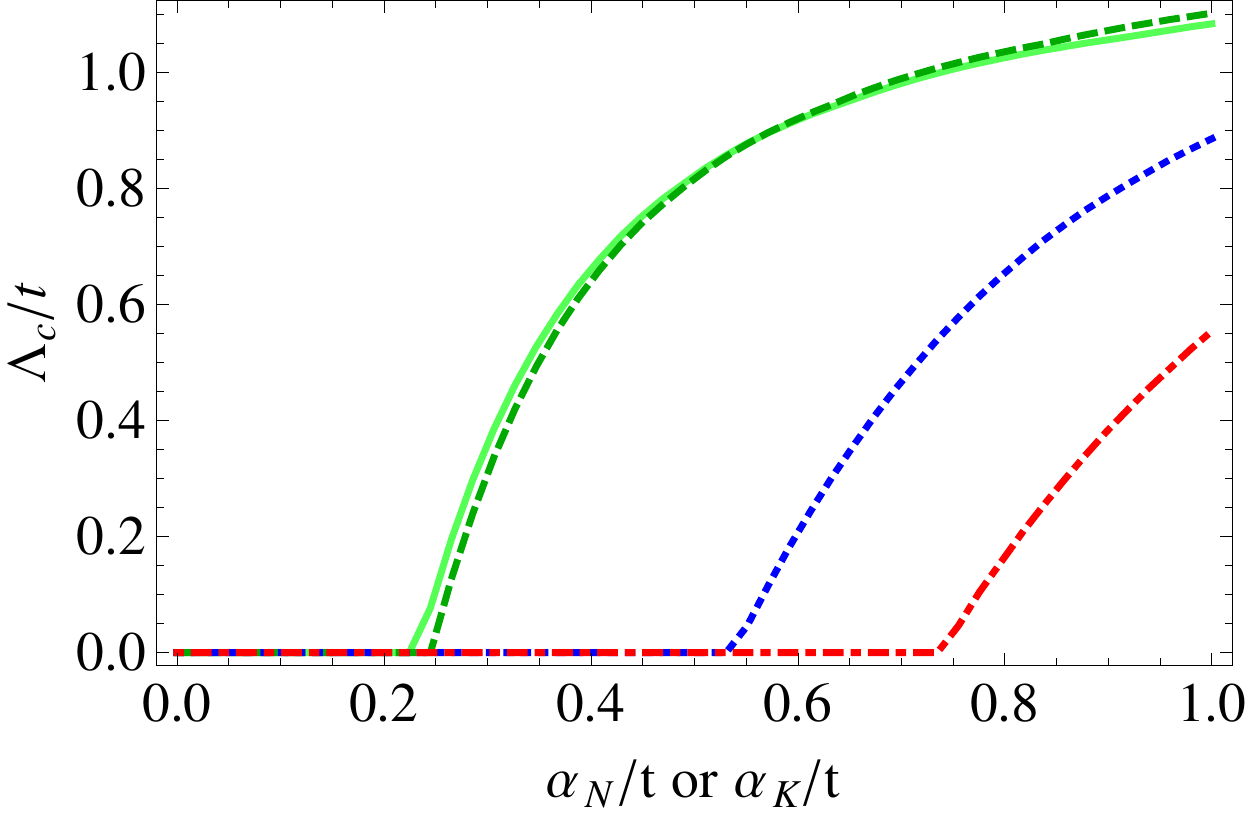}}
\\
\subfigure[Critical scale vs. $\alpha_K/\alpha_N$ for fixed $\alpha_N$. \label{fig:nem_kek}]
{\includegraphics[height=.48\columnwidth]{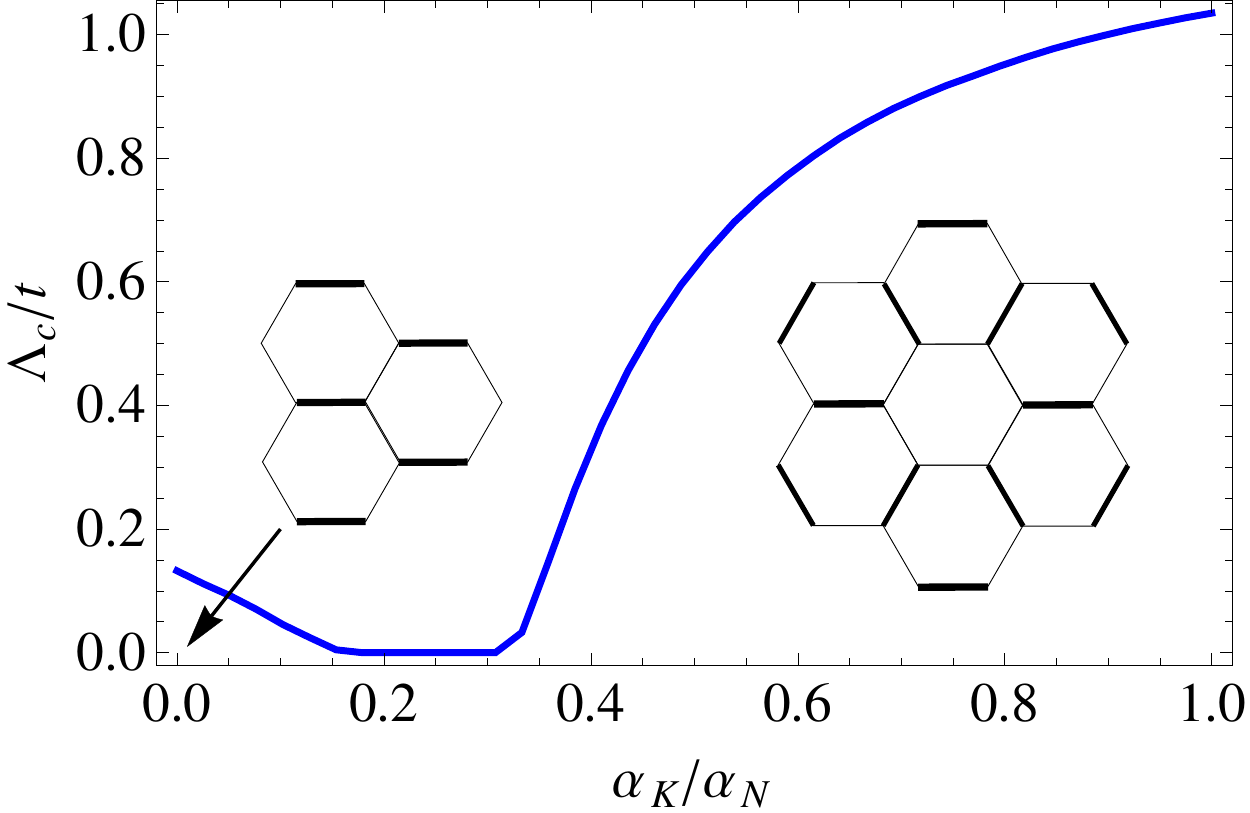}}
\caption{(a) Critical scale as function of the phonon-mediated interaction parameter. For $\alpha_N=\alpha_K$ (dashed) and $\alpha_N=0$ (solid) the Kekul\'e state is induced, whereas for $\alpha_K=0$ (dot-dashed) a nematic bond order develops. With particle-particle contributions only and $\alpha_N=\alpha_K$ (dotted) a conventional superconducting ground state is favoured. \\
(b) Critical scale as function of the ratio $\alpha_K/\alpha_N$ at fixed $\alpha_N=0.8t$. Insets show the nematic ($\alpha_K/\alpha_N \leq0.15$) and the Kekul\'e ($\alpha_K/\alpha_N \geq0.3$) state, respectively.}
\end{figure}

We address several scenarios for the electron-phonon coupling by different choices for the interaction parameters $\alpha_N$ and $\alpha_K$. We study the cases where only phonons from the vicinity of the $\bs{\Gamma}$ point ($\alpha_K=0$), or only phonons close to the Dirac points ($\alpha_N=0$), contribute. Moreover, we include their mutual influence on each other by tuning through different ratios of $\alpha_N/\alpha_K$ with the most physical case around $\alpha_N/\alpha_K\approx1$. The investigated parameter range spans from $\alpha_{N/K}=0$ to $\alpha_{N/K}=t$. In these flows, divergences develop only if the interaction parameter $\alpha_{N/K}$ is large enough, otherwise the system is a stable semimetal. This phenomenon is clearly related to the vanishing density of states at the Fermi level and has been seen for many other types of interaction-driven instabilities for fermions with this spectrum before (e.g. Refs.~\onlinecite{herbut2006,honerkamp2008,raghu2008}). First, we discuss the results for $\alpha_N=\alpha_K$. The discretized, bare interaction, which is the initial value of the flow equation, is shown in Fig.~\ref{fig:start}. Here, the critical parameter value for an instability to occur at half filling is $\alpha_K^c\approx0.28\, t$, cf. Fig.~\ref{fig:alphacrit}. The momentum structure of the effective interaction close to the critical scale is presented in Fig.~\ref{fig:eff_phmed_K}. It has the same structure as the bare phonon-mediated interaction, however, only for momentum transfer with $\mb{k_3-k_1=K}$ and $\mb{k_3-k_1=-K}$. Using this relation in the coupling function $V(\mb{k_1,k_2,k_3})$ gives the effective low-energy Hamiltonian
\begin{widetext}
\al{
H_K&=-\frac{V_{\mathrm{eff}}}{N}\sum_{\lambda}\sum_{\mb{k},s}\rund{g_{\mb{k}}^{\lambda}(\mb{K})c_{A,\mb{k},s}^{\dagger}c_{B,\mb{k-\mb{K}},s}+g_{\mb{k}}^{\lambda}(\mb{-K})^*c_{B,\mb{k+K},s}^{\dagger}c_{A,\mb{k},s}}\nonumber\\
 &\hspace{2.5cm}\times\sum_{\pr{\mb{k}},\pr{s}}\rund{g_{\pr{\mb{k}}}^{\lambda}(\mb{-K})c_{A,\pr{\mb{k}},\pr{s}}^{\dagger}c_{B,\pr{\mb{k}}+\mb{K},\pr{s}}+g_{\pr{\mb{k}}}^{\lambda}(\mb{K})^*c_{B,\pr{\mb{k}}-\mb{K},\pr{s}}^{\dagger}c_{A,\pr{\mb{k}},\pr{s}}}
}
with $V_{\mathrm{eff}}>0$. For this expression, we can perform a mean-field decoupling with the molecular field $\Delta_{\mathrm{ph}}^{\lambda}(\mb{Q})=\frac{V_{\mathrm{eff}}}{N}\sum_{\mb{k},s}\spitz{g_{\mb{k}}^{\lambda}(\mb{K})c_{A,\mb{k},s}^{\dagger}c_{B,\mb{k-K},s}+g_{\mb{k}}^{\lambda}(-\mb{K})^*c_{B,\mb{k+K},s}^{\dagger}c_{A,\mb{k},s}}$ yielding
\al{\label{MF-Kekule}
H_{K}&\approx-\sum_{\mb{k},s,\lambda}\left[\Delta_{\mathrm{ph}}^{\lambda}(-\mb{K})\rund{g_{\mb{k}}^{\lambda}(\mb{K})c_{A,\mb{k},s}^{\dagger}c_{B,\mb{k-K},s}+g_{\mb{k}}^{\lambda}(-\mb{K})^*c_{B,\mb{k+K},s}^{\dagger}c_{A,\mb{k},s}}\right. \notag\\
 &\hspace{2.5cm}+ \left.\Delta_{\mathrm{ph}}^{\lambda}(\mb{K})\rund{g_{\mb{k}}^{\lambda}(-\mb{K})c_{A,\mb{k},s}^{\dagger}c_{B,\mb{k+K},s}+g_{\mb{k}}^{\lambda}(\mb{K})^*c_{B,\mb{k-K},s}^{\dagger}c_{A,\mb{k},s}}\right],
}
ignoring the constant term.
\end{widetext}

If we compare this expression to the coupling Hamiltonian, Eq.~(\ref{eq:eph_coupling}), we recover the contribution of the $\pm\mb{K}$-phonons from the beginning with the identification $\Delta_{\mathrm{ph}}^{\lambda}(\pm\mb{K})=\abs{\alpha_{||}}u_{\pm\mb{K}}^{\lambda}$. But in order to get a diverging phonon-mediated interaction parameter $\alpha_K$, the phonon frequency must tend to zero. Thus the observed instability results in a static lattice distortion formed by the modes from $\pm\mb{K}$, which are responsible for the Kekul\'e distortion shown in the inset of Fig.~\ref{fig:nem_kek}. Correspondingly, close to the Dirac points, the eigenenergies extracted from Eq.~(\ref{MF-Kekule}) coincide with previous investigations of the Kekul\'e phase\cite{Chamon1999,Hou2007}. In this state, the hopping is non-uniform in the three bond directions resulting in a tripled unit cell and the opening of a gap.

Now we tune the interaction to both extreme cases where one of the parameters is zero. For $\alpha_N=0$, we again find the Kekul\'e distortion as leading instability. The only difference is that the critical $\alpha_K$ needed to induce the ordering is slightly decreased to $\alpha_K^c\approx0.26t$, as visible in Fig. \ref{fig:alphacrit}. This already shows that the small-wavector phonons controlled by $\alpha_N$ have a destructive influence on the large-wavevector phonons controlled by $\alpha_K$, i.e. there is some degree of phonon-phonon interaction.
  
For $\alpha_K=0$, the behavior is qualitatively different. As shown in Fig. \ref{fig:alphacrit}, the instability does not occur until a threshold value $\alpha_N^c\approx0.78t$ is reached and cannot be due to $\pm\mb{K}$-phonons because they are not included in this case. Instead we find the momentum structure at low energies of Fig.~\ref{fig:eff_phmed_N}, which mirrors the bare phonon-mediated interaction for zero momentum transfer $\mb{k_3=k_1}$. The extracted, low-energy Hamiltonian is
\al{
  H_{N}&=-\frac{\bar{V}_{\mathrm{eff}}}{N}\sum_{\lambda}\sum_{\mb{k},s}\rund{g_{\mb{k}}^{\lambda}c_{A,\mb{k},s}^{\dagger}c_{B,\mb{k},s}+g_{\mb{k}}^{\lambda*}c_{B,\mb{k},s}^{\dagger}c_{A,\mb{k},s}}\nonumber\\ 
  &\times\sum_{\pr{\mb{k}},\pr{s}}\rund{g_{\pr{\mb{k}}}^{\lambda}c_{A,\pr{\mb{k}},\pr{s}}^{\dagger}c_{B,\pr{\mb{k}},\pr{s}}+g_{\pr{\mb{k}}}^{\lambda*}c_{B,\pr{\mb{k}},\pr{s}}^{\dagger}c_{A,\pr{\mb{k}},\pr{s}}}
}
with $\bar{V}_{\mathrm{eff}}>0$ and the abbreviation $g_{\mb{k}}^{\lambda}(\mb{q}=0)=g_{\mb{k}}^{\lambda}$. The corresponding mean-field Hamiltonian results in
\begin{equation}
H_{N}\approx-2\sum_{\mb{k},s,\lambda}\Delta_{\mathrm{ph}}^{\lambda}\rund{g_{\mb{k}}^{\lambda}c_{A,\mb{k},s}^{\dagger}c_{B,\mb{k},s}+g_{\mb{k}}^{\lambda*}c_{B,\mb{k},s}^{\dagger}c_{A,\mb{k},s}},
\end{equation}
where $\Delta_{\mathrm{ph}}^{\lambda}=\frac{\bar{V}_{\mathrm{eff}}}{N}\sum_{\mb{k},s}\spitz{g_{\mb{k}}^{\lambda}c_{A,\mb{k},s}^{\dagger}c_{B,\mb{k},s}+g_{\mb{k}}^{\lambda*}c_{B,\mb{k},s}^{\dagger}c_{A,\mb{k},s}}$ is the molecular field .
Comparison to the coupling Hamiltonian, Eq.~(\ref{eq:eph_coupling}) now gives the static lattice distortion due to the zone center $E_2$ phonons with $2\Delta_{\mathrm{ph}}^{\lambda}=\abs{\alpha_{||}}u_{\mb{0}}^{\lambda}$. For $\mb{q}=0$,  neighboring sites have different signs $\mb{e}_{\mb{0},A}^{\lambda}=-\mb{e}_{\mb{0},B}^{\lambda}$. This means that the two sub-lattices are moved in opposite directions in this state.
The sixfold symmetry of the original lattice is reduced to a twofold one, the translational symmetry of the underlying Bravais lattice, however, is maintained, corresponding to a nematic ordering pattern. The best energy gain is a distortion along the bonds between two sites. As a result, we obtain the configuration shown in the inset of Fig.~\ref{fig:nem_kek}, where the hopping along one bond direction is enhanced and the Dirac points are shifted away from the Brillouin zone corners. Such a state was studied in Ref.~\onlinecite{Chamon1999}.

In Fig. \ref{fig:nem_kek}, we also show the evolution of the critical scale for fixed supercritical $\alpha_N$ when $\alpha_K$ is increased. We clearly see that the 'Kekul\'e phonons' with large wavevector transfer reduce the scale for the nematic instability and push it to zero already for $\alpha_K \approx 0.2 \alpha_N$. This exhibits a clear 'anharmonic' interaction between phonon modes with different wavevectors that is revealed by the fRG treatment. When $\alpha_K$ is increased further, one reaches the Kekul\'e-ordered phase again. Note that the rather high critical scales found here are not to be taken literally, due to the mentioned overestimation of the phonon effects when the retardation is ignored.

Usually, in more than one dimension, an important property of the electron-phonon interaction is to induce Cooper pairing, which seems to be suppressed here. We can indeed recover a conventional phonon-mediated superconducting state, but only if the RG flow equation for the interaction is reduced to the particle-particle term and all particle-hole terms are switched off. Through this the integration of the fRG equations is identical to a ladder summation in the particle-particle channel. However, the critical interaction strength needed to observe a flow to strong coupling for such an undisturbed Cooper instability is larger than in the bond-odering case. Here, without the particle-hole term, we find $\alpha^{SC}\approx0.58t$ for $\alpha_K=\alpha_N$. This shows that phonon-mediated superconductivity arises in the particle-particle channel as one would expect, however only when the competing contributions from the particle-hole channels are completely neglected.


\subsection{Inclusion of density-density interactions}\label{coulomb}

%
\begin{figure}[t]
  \subfigure{\includegraphics[width=0.4\textwidth]{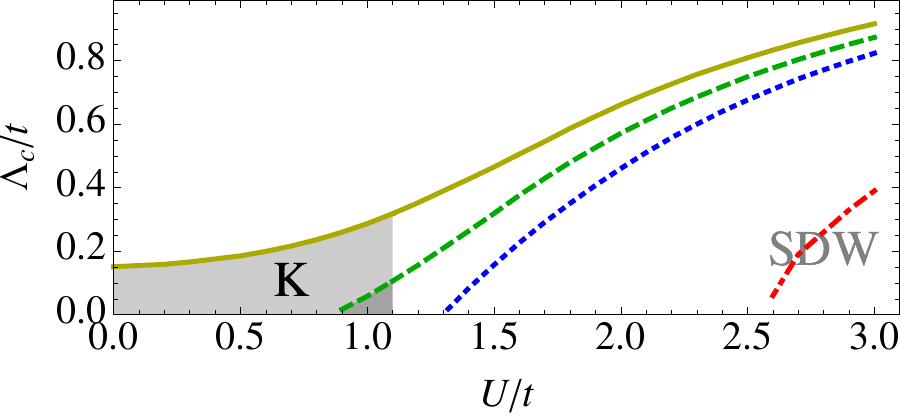}}
\subfigure{\includegraphics[width=0.4\textwidth]{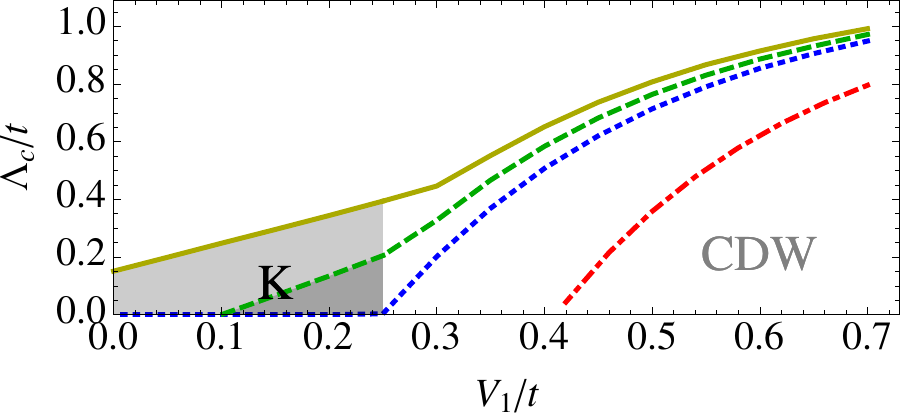}}
  \subfigure{\includegraphics[width=0.4\textwidth]{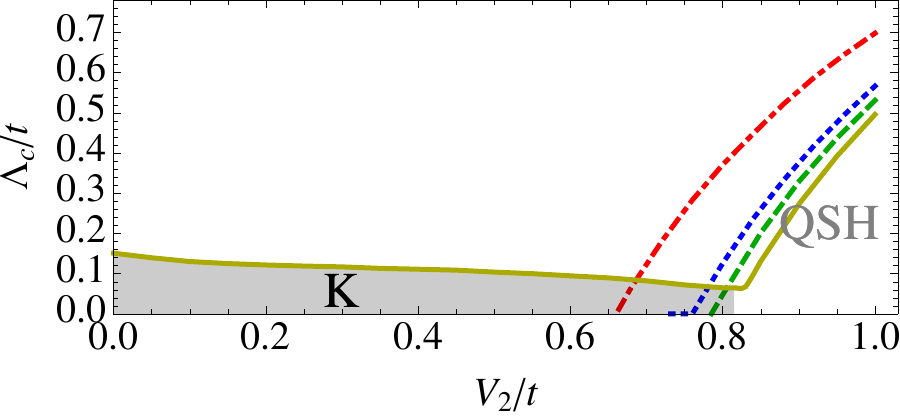}}
  \caption{Critical scale as function of the density-density interaction $U, V_1$ and $V_2$ (from top to bottom panel) for different values of the ph-med interaction $\alpha\in\{0.28t,0.24t,0.20t,0\}$ (solid, dashed, dotted, dot-dashed). For small density-density interactions the gray-shaded area shows the appearance of the Kekul\'e ordering tendency (K). For larger density-density interaction we recover SDW (onsite interaction, top panel), CDW (nearset-neighbor i.a., middle panel) and QSH (next-to-nearest neighbor i.a., bottom panel)}
  \label{fig:critU}
\end{figure}

We now also include the Coulomb-induced repulsive density-density interactions $U$, $V_1$ and $V_2$ as given in Eq.~(\ref{eqn:interactions}). First, we include, in addition to the phonon-mediated interaction, each one of the three short-ranged interactions $U$, $V_1$ and $V_2$ separately. This shows if the phonon-mediated interaction amplifies or weakens the effect of the respective electronic interaction. The results are compared to the case without the consideration of phonons.

Running the fRG flow with a fixed, supercritical on-site interaction for different phonon-mediated interaction strengths leads to an antiferromagnetic spin density wave (SDW) as in the case without phonons. But with increasing phonon-mediated interactions, the critical scale of the flow is enhanced. This amplifying tendency is also observed if we determine the critical on-site interaction needed to induce an instability. It reduces from $U^c=2.6t$ to $U^c=1.3t$ if a phonon-mediated interaction of, e.g. $\alpha=0.2t$, is turned on, cf.~Fig.~\ref{fig:critU}. For the nearest-neighbor interaction, we obtain qualitatively the same behavior. As without phonons, the nearest-neighbor interaction triggers a charge density wave (CDW) whose critical scale is increased with increasing phonon mediated interaction. However, this effect is not as large as in the case of the on-site interaction. Nevertheless, the critical $V_1$ changes from $V_1^c=0.4t$ for $\alpha=0$ to $V_1^c=0.25t$ for $\alpha=0.2t$. The situation is different if we consider only a next-nearest-neighbor interaction which induces a quantum spin Hall state (QSH). Including an electron-phonon coupling suppresses the tendency for the formation of a QSH state as shown in the lower panel of Fig.~\ref{fig:critU}.

\begin{figure}[t]
  \centering
  \includegraphics[width=.93\columnwidth]{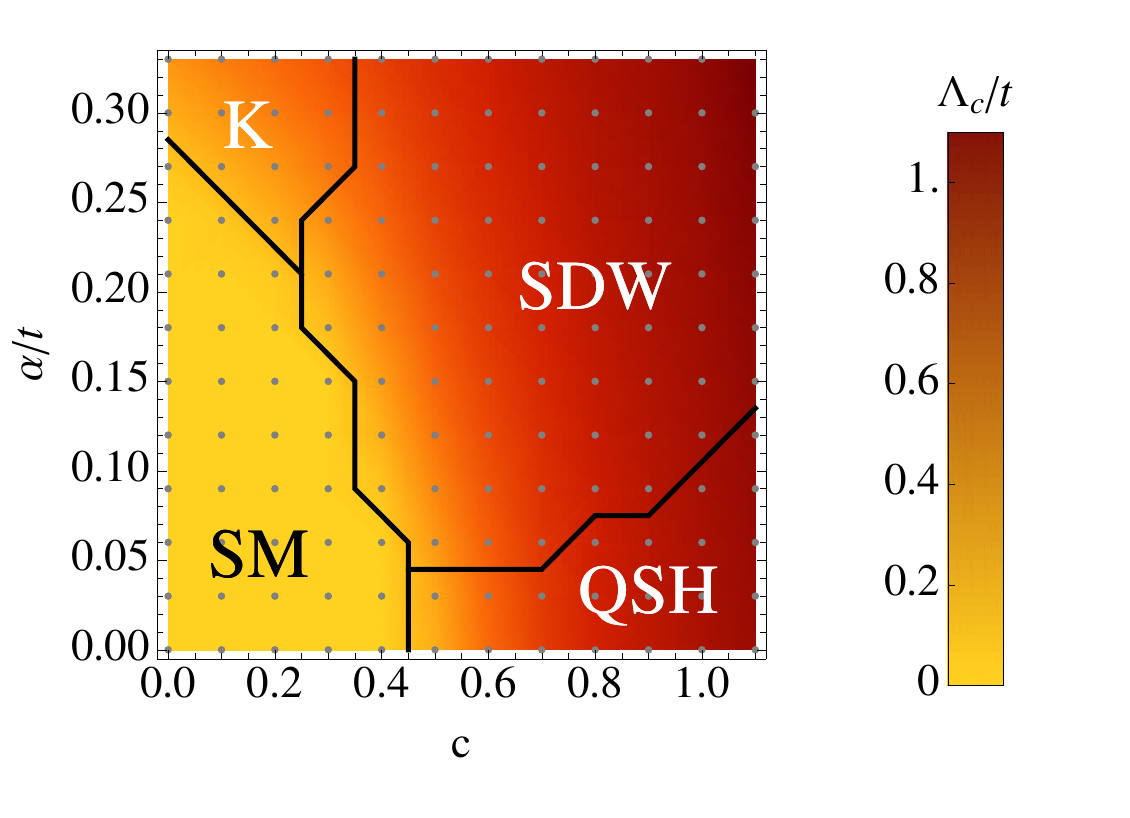}
  \caption{Phase diagram for the rescaled ab initio density-density interaction profile in graphene from Ref.~\onlinecite{wehling2011} with rescaling parameter $c$ and the electron-phonon coupling strength $\alpha=\alpha_N=\alpha_K$. The density-density interactions are rescaled according to $\{U/t,V_1/t,V_2/t\}\approx \{3.3,2.0,1.5\}\rightarrow c\{U/t,V_1/t,V_2/t\}$. In case without electron-phonon coupling we find a Quantum Spin Hall state being favored for this interaction profile, cf. Ref.~\onlinecite{Scherers2012}. In agreement with the previous observations the EPC supports the tendency towards the SDW phase.
\label{fig:phasediag}}
\end{figure}

These tendencies are confirmed when we run the fRG flow with all interactions, i.e.~density-density repulsion up to the second nearest neighbor and the phonon-mediated interaction, included. The parameter range which we account for extends from zero through the cRPA values from Ref.~\onlinecite{wehling2011}. They are taken as upper bounds because the fRG tends to overestimate the critical scales. A summary of this investigation is given by Fig.~\ref{fig:phasediag}.
%

\section{Conclusions}\label{conclusions}

In this work we have analyzed the impact of in-plane phonons on possible ground state orderings in a simple theoretical model for monolayer graphene. We focused on phonon eigenmodes arising from the modulation of nearest-neighbor bonds between the carbon $\pi$-orbitals on the honeycomb lattice with wave vectors near $\bs{\Gamma}$ and $\mb{K}/-\mb{K}$. These modes, classified as $E_2$ for small wavevector and $A'_1$, $A_1$, $B_1$ for large wavevector transfer,  are known from DFT calculations to couple most strongly to the electrons. 

The electron-phonon coupling and the phonon dynamics for these modes was transformed into an effective electron-electron interaction, with some idealizations. Studying the effects of the phonon-mediated interaction without including Coulomb interactions between the electrons we find the following results. Near charge neutrality (i.e. for the undoped system) the dominant instability is in the particle-hole channel, and not in the pairing channel as is usually the case in non-nested systems. While our study may not be fully quantitative, the picture we find is that the phonons with large momentum transfer dominate in the low-energy effective interactions and that the predominant instability is towards Kekul\'e bond order, where the unit cell is tripled by a pattern of strengthened and weakened bonds. This state opens a gap at the Dirac points, i.e. is an insulator. Various works have argued for the existence of this state due to Coulomb interactions\cite{nomura2009,kharitonov2012}. Previous RG studies of the same model \cite{herbut2006,honerkamp2008,Scherers2012} did not find the Kekul\'e order for Coulomb interactions of density-density type in the effective model, but it now occurs due to the bond-bond interactions mediated by the phonons. We can also weaken the influence of the large-wavevector phonons in the effective electron-electron interaction, emphasizing the small-$q$ phonons. In this case, a nematic instability becomes dominant where one of the three bond directions is enhanced with respect to the other two directions. The resulting spectrum features shifted Dirac points. Considering the  competition between the different phonon channels, we found that the large-wavevector Kekul\'e phonons considerably weaken the tendencies towards nematic order driven by the small-wavevector phonons. This means that there is a significant amount of non-RPA  or anharmonic physics at low energy scales, where phonons with different wavevectors interact destructively.

More realistically, the phonon-mediated interaction should be considered together with the Coulomb interaction between the electrons. The Coulomb interactions alone have been studied with RG and many other methods on honeycomb lattices in a number of works (e.g. Refs.~\onlinecite{tchougreff1992,herbut2006,honerkamp2008,raghu2008}).  In particular, quantum Monte Carlo calculations\cite{sorella1992,meng,sorella2012} have firmly established  that the ground state for pure onsite repulsions becomes an antiferromagnetic spin-density-wave (SDW) state when the Hubbard-$U$ exceeds a threshold value. For  interactions that extend further in space, only less controlled techniques are applicable. RG and saddle-point calculations\cite{raghu2008} found that charge-density wave states and interaction-induced quantum spin Hall (QSH) states are relevant competitors, depending on the profile of the effective interaction. Adding phonons to this interplay of the electronic ordering tendencies shifts the balance toward the SDW, while the competing QSH channel gets weakened. Interestingly, the bond phonons considered in the work actually increase the SDW and also potential CDW ordering tendencies. This can be seen most clearly in the lowering of the threshold value for the Hubbard interaction $U$ to change the semi-metal into the SDW state when the electron-phonon interaction is turned on. For the QSH state we found the reversed trend, indicating a destructive interplay with the phonons. 

Hence, one important upshot of our study is the identification of the most relevant phonon-mediated effects on the ground state. Based on our study, we do not expect that the nature of potential ground state ordering or, more realistically according to the experimental state in single-layer graphene, the nature of the leading correlations if the overall interaction strength is insufficient to gap the semimetal, is determined by a phonon-mediated instability. However, we have shown that the phonon sector may actually shift the phase boundaries between different electronically-driven ordering tendencies. Hence, phonon effects may yet play an important role in deciding which of these channels wins. We can also try to extrapolate our results for the single-layer honeycomb lattice to multi-layer graphene, where experiments indeed show that the semimetal gives way to a gapped state at lower temperatures\cite{bao}. From our theoretical experience with multi-layer honeycomb systems\cite{Scherers2012} we can state that the main ordering tendencies are still the ones of the single layer, while the stacking only modifies the available density of states at low scales. Then, we should expect that in the multi-layer system, the phonon degrees of freedom have a similar influence in modifying the interplay of the electronic ordering tendencies as we find here. This makes the SDW state a more robust candidate to explain the observed gaps. Notably, even though the SDW state does not feature spin-resolved edge states like the QSH state, it may still be useful resource for nano-spintronic devices when the multilayer system is slightly doped and gated\cite{yuan}.

{\bf{\it{Acknowledgements}}}\\
%
We acknowledge discussions with L.~Boeri and S.A. Maier. M.M.S. is supported by the grant ERC-AdG-290623 and DFG grant FOR 723. C.H. acknowledges support from DFG FOR 912 and SPP 1459. 


\end{document}